\documentstyle[11pt,aasms]{article}

\def\mass{${\cal M}$}
\def\hst{{\it HST\/}}
\def\etal{{\it et al.\/}}
\def\msun{${\cal M}_\odot$}

\def\minspt{$\buildrel{\prime}\over .$}

\def\Vsix{$V_{606}$}
\def\Vfive{$V_{555}$}
\def\Ieight{$I_{814}$}
\def\V6{$V_{606}$}
\def\V5{$V_{555}$}
\def\I8{$I_{814}$}
\def\MV{$M_{555}$}
\def\MI{$M_{814}$}

\def\hst{{\it HST\/}}
\def\HST{{\it HST\/}}
\def\simless{$\buildrel{<}\over \sim$}
\def\simgr{$\buildrel{>}\over \sim$}

\begin{document}

\title{A Comparison of Deep HST Luminosity Functions\\of Four Globular
Clusters\altaffilmark{1}}

\author{Giampaolo Piotto\altaffilmark{2}, Adrienne M.~Cool\altaffilmark{3}, 
and Ivan R.~King\altaffilmark{4}}

\altaffiltext{1}{Based on observations with the NASA/ESA {\it Hubble Space Telescope},
obtained at the Space Telescope Science Institute, which is operated by AURA, Inc., under 
NASA contract NAS5-26555}

\altaffiltext{2}{Dipartimento di Astronomia, Universit\` a di Padova, Vicolo dell'Osservatorio 5, I-35122 Padova, Italy}

\altaffiltext{3}{Astronomy Department, University of California,
Berkeley, CA 94720-3411, and Department of Physics and Astronomy, San
Francisco State University, 1600 Holloway Avenue, San Francisco, CA 94132}

\altaffiltext{4}{Astronomy Department, University of California, Berkeley, CA 94720-3411}

\begin{abstract}
From deep color--magnitude arrays made from $V$ and $I$ images taken
with {\it Hubble Space Telescope\/}'s WFPC2 camera we have determined
luminosity functions (LFs) down to a level that corresponds to
$\sim$0.13 \msun, for the low-metal-abundance globular clusters M15,
M30, M92, and NGC 6397.  Because of the similarity of the metallicities
of these clusters, differences in their luminosity functions directly
trace differences in their mass functions.  The LFs of M15, M30, and M92
agree closely over the entire observed range, whereas that of NGC 6397
drops away sharply at the faintest magnitudes.  We suggest that the
deficiency of low-mass stars in NGC 6397 is due to tidal shocks, to
ejection through internal relaxation, or to a combination of the two.
With the presently available mass--luminosity relations, we find that
even in M15, M30, and M92 the mass functions probably do not rise so
fast as to make the low-mass end dominant.
\end{abstract}


\section{Introduction}

The {\it Hubble Space Telescope} (\hst) now makes it possible to
derive color--magnitude diagrams (CMDs) of globular clusters (GCs)
that extend several magnitudes fainter than the CMDs found for the
same clusters from the ground.  Among the important by-products of
these CMDs are luminosity functions (LFs), and from them mass
functions (MFs), that can extend over nearly the entire length of the
lower main sequence.

Insight into the formation and dynamical evolution of globulars can be
gained by using these LFs and MFs to compare the stellar content of
different clusters.  It has often been suggested that the mass
function of a cluster depends on its position in the Galaxy, its
metallicity, and its dynamical history (see, e.g., McClure \etal\
v1986, Aguilar, Hut, \& Ostriker 1989, Djorgovski, Piotto, \&
Capaccioli 1993); and the relationships are complicated by the fact
that the latter two factors correlate with the first.  In any case,
the main-sequence mass functions of clusters are one of our important
clues relating to their origin.

Furthermore, globular clusters have quite often been used as tracers and
indicators of the halo population of which they are the most conspicuous
part.  In particular, it has been suggested that there is a steep rise
at the low-mass end of cluster MFs (Richer \etal\ 1991), and that this
might be indicative of a considerable contribution by low-mass stars to
the little-known total mass that governs the poorly understood flat
rotation curve of the Galaxy (Richer \& Fahlman 1992).
In the case of the Galaxy, the steep MF slope that Richer and 
Fahlman (1992) found at the faint end has been contradicted by 
better data by Reid \etal\ (1996); and we shall show here that their
assertion about steep MFs for low-mass stars in globular clusters also
fails to be borne out by improved data that we will present below.  Of equal
importance will be the demonstration that with one exception that is
probably understandable, our sample of clusters shows a remarkable
uniformity of mass functions.

Progress on these issues requires accurate photometry of low-mass
main-sequence stars.  Ground-based studies have so far been limited
largely to the mass range above 0.4\msun; deeper studies have been
pursued only with difficulty and uncertainty.  With the advent of
\hst, however, we can largely circumvent the crowding problems that 
plague ground-based observations, reaching several magnitudes deeper,
down to masses close to the bottom end of the 
hydrogen-burning main sequence.

Here we present the first deep \hst\ LFs of M30 (NGC 7099) and M92
(NGC 6341), and independent measurements of the LFs of NGC 6397 and
M15 (NGC 7078). The CMD and the LF of NGC 6397 have already been
presented in Cool, Piotto, \& King (1996, CPK).  We compare our
results for the latter two clusters with \hst\ LFs of Paresce, De
Marchi, \& Romaniello (1995) and De Marchi \& Paresce (1995), 
and then compare the four cluster LFs with each other.  The results of our
comparison differ from those reached by De Marchi \&\ Paresce (1995),
in that we find a significant difference between the LFs of NGC 6397
and M15.

Two features of this set of four clusters make for a particularly
useful comparison.  First, three of them have collapsed cores and
similar surface-brightness profiles, which make the conversion from
the local, observed LFs to global LFs straightforward.  M92 is a
high-concentration cluster; in this case also the observed LF does not
differ appreciably from the global one.  Second, all four clusters
have comparable metallicities (see Table 1), so that similarities or
differences in their LFs will directly reflect similarities or
differences in their MFs.

A description of the observations and of the methods of analysis are
given in Section 2.  We present the cluster LFs and MFs in turn in
Sections 3 and 4, exploring in the latter the role of the
mass--luminosity relation.  The findings are summarized and discussed
in Section 5.

\section{Observations and Analysis}

The data were taken with the WFPC2 in parallel mode, in fields
$\sim$4\minspt6 from the center of each cluster.  
(More exactly, the
centers of chips W2, W3, and W4 were respectively 5.46, 4.28, and 3.90
arcmin from the center of each cluster, the FOC having been pointed at
the cluster center in every case.)  The observation dates and total
exposure times are given in Table 1.  The NGC 6397 and M30 exposures
were obtained with the F814W and F555W filters, and are part of our
Cycle-4 observing program.  We included short ($\sim$60s) exposures in
this program, which allow us to measure stars that were saturated in the
long ($\sim$1000s) exposures.  The M92 images are part of our Cycle-5
program and were taken with the F814W and F606W filters; short (30s,
100s) exposures were included, making it possible to measure stars up to
the turnoff.  The choice of the F606W filter, broader than the F555W
filter, enabled us to obtain images of nearly matching depth in the two
bands with equal exposure times.  The M15 exposures, by contrast, came
from a generic parallel-exposure program that used the F814W and F606W
filters and lacked the short exposures.  They are the same images
analyzed by De Marchi \&\ Paresce (1995).

Each single long-exposure image was cleaned of cosmic rays using the
method described by Anderson, King, \& Sosin (1995).  
Images with identical pointings
were averaged.  Lists of stars were created for each of the
F814W averaged images using DAOPHOT/FIND (Stetson 1987, 1991),
modified as described in CPK to help eliminate false detections along
diffraction spikes, in the halos of bright stars, and at the positions
of hot pixels.  The removal of as many of these artifacts as possible
was essential for a reliable LF.  As in all cases the F814W images
were the deepest images, we used the F814W star list for the F555W and
F606W images.  We measured the magnitudes of the stars using the
hybrid weighted, neighbor-subtracted aperture photometry technique
described by Cool \& King (1995).  The first step was to construct a
model PSF for each image (using from 100 to 200 stars), consisting of
a Moffat function plus residuals, allowing quadratic variation of the
residuals with position in the frame.  PSF-fitting results from
ALLSTAR were then used to subtract all the stars from each
image. Next, stars were added back into the ``subtracted'' image one
at a time, reproducing the original data exactly for the added star,
but with all the neighbors removed. The flux of each
``neighbor-subtracted'' star was then measured in an aperture, with
the local background estimated using the DAOPHOT routine (Stetson
1987). The aperture magnitude was obtained by weighting each pixel by
the reciprocal of its expected variance (Cool \&\ King 1995). The
weighted, neighbor-subtracted aperture photometry produced a
noticeably tighter main sequence at the faint end than did DAOPHOT
used in the conventional way, and the improvement in photometric
accuracy was further confirmed in artificial-star experiments.

We transformed the F814W and F555W instrumental magnitudes into the
WFPC2 ``ground system'' using Eq.\ 6 of Holtzman \etal\ (1995) and 
the coefficients in their Table 6.  We adopt this photometric system where
possible, as it has become the \HST\ standard.  Hereafter we will
refer to these calibrated magnitudes as \I8\ and \V5.  For the range
of colors sampled here, the \I8\ and \V5\ magnitudes should match $V$
and $I$ magnitudes in the Johnson/Cousins system to within a few
hundredths of a magnitude (see Fig.\ 8 in Holtzman \etal).  As the
WFPC2 ground system is not defined for the ``wide V'' (F606W) filter
used to image M92 and M15, we transformed these instrumental
magnitudes directly into the ``standard'' (Johnson/Cousins) system
instead, using Eq.\ 9 and the coefficients in Table 10 of Holtzman
\etal\ (1995).  Hereafter we refer to these calibrated magnitudes as
\Vsix.  For the range of colors sampled in our data, \Vsix\ should
match Johnson $V$ to within a few hundredths of a magnitude (see Fig.\
10 in Holtzman \etal).

In the case of M15 all exposures were aligned, so that we had a single
averaged image to work with in each filter.  In the case of the other
three clusters the fields were not perfectly overlapping and aligned:\
for NGC 6397 we had 5 averaged images per filter; the number was 3 for
M30, and 2 for M92.  For each cluster we generated a master list of
stars by computing transformations between frames, first correcting for
geometrical distortion in the chips, and then requiring that positions
agree to within 1.5 pixels (a matching criterion loose enough to avoid
selective loss of faint stars). Final magnitudes were derived by
flux-averaging the 2 (M92) to 5 (NGC 6397, in the overlapping regions)
independent measurements in each filter, so that the full benefit of the
information contained in the complete data set was recovered.  Note that
the CMDs of NGC~6397, M15, and M30 presented in this paper come from the
three WF chips combined; the CMD of M92 comes from the WF4 chip alone.
The LFs have been obtained from the 3 WF chips for NGC~6397, from chip
WF3 for M15 and M30, and from chip WF4 for M92.  Adding the data from
the other chips would marginally increase the statistical significance
of the LFs.  In view of the small error bars of the present LFs (see
Fig. 2), we considered it not worth the large cpu time that would have
been required to run the crowding experiments for every chip (see
below).

In Figure 1 are the color--magnitude diagrams derived from this
photometry.  The diagram for NGC 6397 has been presented by CPK and is
shown here for purposes of comparison.  Detailed discussions of the
other CMDs will appear elsewhere.  Here we focus on the main-sequence
luminosity functions derived from these CMDs.  In all four cases we
measure stars from the turnoff (or just below it) to a limiting
magnitude \Vfive,\Vsix\ \simgr\ 27.  The main sequence (MS) is well
defined, spanning from 8 (M15) to nearly 10 magnitudes in \Vfive\
or \Vsix. All four MSs show characteristic bends, which are
particularly important for best tuning the theoretical models
(Alexander \etal\ 1996, and Section 4 of this paper).  For
\Vfive,\Vsix\ $>$ 25, measurement errors cause the MS to broaden
noticeably, but its ridge can be distinguished down to limiting
magnitudes that correspond to an absolute magnitudes in the range
\MV\ $\sim$ 12--13 
for M15, M30, and M92 (see adopted distance moduli
in Table 1).  Only in the case of NGC 6397 does field contamination
prevent a reliable identification of the faintest MS stars, below
\Vfive $\sim26$, which still correponds to an absolute magnitude
\MV $\sim14$ in this cluster.

Below and to the left of the main sequence in NGC 6397 is the cluster's
white dwarf cooling sequence (see CPK for a detailed description and
analysis).  No obvious white dwarfs are identifiable in the other three
clusters (apart from one object at $V=24.6$ and $V-I=-0.05$ in
M15)---which is to be expected, since all have apparent distance moduli
that are at least 2.3 magnitudes greater than NGC 6397.

Contamination by background and foreground stars is negligible for
M15, M30, and M92, as expected from their galactic latitudes
($|b|>27^\circ$).  By contrast, a large number of objects to the left
and the right of the MS of NGC 6397 are background halo and foreground
disk stars, respectively, the latter due to the cluster's low latitude
($b=-12^\circ$). Their numbers are in reasonable agreement with those
predicted by the Bahcall \& Soneira (1980) Galaxy model. In all four
diagrams, at the faintest magnitudes there is probably also an
admixture of unresolved galaxies.

Particular attention was devoted to determining the completeness of
our samples.  For each cluster we created 40 to 45 new images, using
DAOPHOT (Stetson 1987, 1991).  Each of these was the original image
with up to 200 artificial stars added (no more, lest we alter the
degree of crowding).  Each artificial star was a replica of the PSF,
with Poisson noise added; the star was given a main-sequence color,
and geometrical transformations were used to place it at the same
randomly chosen position in both the $V$ and the $I$ image.  The
images containing artificial stars were measured in the same way as
the original images.  In the LFs presented here, we include only
points for which the completeness figures were 50\% or higher,
so that none of the counts have been corrected by more than a factor of 2.  
For NGC 6397 and M30, the completeness is everywhere greater than 80\%.
The completeness drops below this level only in the last point of the M15 
LF and in the last 3 points of the M92 LF.  

The use of
color--magnitude diagrams allows us to distinguish cluster stars from
field stars.  In three of the four clusters field contamination is
very small, and this is an easy task:\ for M15, M30, and M92 the LF
was obtained by counting stars within $\pm3$ sigmas of the MS ridge
line, and otherwise ignoring field-star contamination.  By contrast,
Fig.\ 1 shows that the CMD of NGC 6397 suffers much interference from
background and foreground stars.  In this case we drew a main-sequence
ridge line, read off its colors, and subtracted the corresponding
main-sequence color from that of each star, so as to create a
verticalized main sequence.  
As described in detail by CPK, we then drew lines on either side of
the MS that excluded field stars, and corrected the regions between
the lines for field-star contamination by using the star density in 
the neighboring regions outside them.

\section{The Luminosity Functions}

The LFs for the 4 clusters are shown in Fig.\ 2 and listed in Tables
2--5 with the completeness estimates.  In Table 2--5, Col. 1 gives the
\Vfive\ or \Vsix\ magnitudes, Col. 2 gives the completeness in the
corresponding magnitude bin, and Col. 3 lists the actual counts,
corrected for completeness and field contamination (where needed). The
same figures for the \Ieight\ LFs are in Cols. 4--6.  Fig.\ 2a shows the
LFs from the \Ieight\ photometry, while Fig.\ 2b shows the \Vfive\ and
\Vsix\ LFs; as the \Vfive\ and \Vsix\ magnitudes should differ by at
most a few hundredths of a magnitude in the color range covered by the
CMDs in Fig.\ 1, a direct comparison of the $V$ LFs is
permissible. Adopted distance moduli are given in Table 1. The distance
moduli and reddenings have been taken from the following papers:\
Piotto, Ortolani, \& Zoccali (1996) for NGC 6397; Durrell \& Harris
(1993) for M15; Piotto \etal\ (1990) for M30; Stetson \&\ Harris (1988)
for M92.  For NGC 6397, stars fainter than the turnoff were saturated
even in our 60-sec exposures, so that the \HST-based LF begins at
\Ieight\ $\sim16.5$ (\Vfive\ $\sim17.1$).  For this cluster, we extended
the LFs upward to the turnoff using ground-based LFs obtained at a
similar distance from the cluster center; for the \Ieight\ LF we used
data from Drukier \etal\ (1993) and for the \Vfive\ LF data from Piotto
\etal\ (1996).  In both cases the \hst\ and ground-based
LFs agree well in the overlap regions (compare the filled triangles
and open squares in Fig.\ 2a,b).

In the absence of a means of normalizing the four LFs to a global
cluster parameter, arbitrary constants determine the vertical
positioning of the individual LFs in Figs.\ 2a and 2b.  We have chosen
these constants in such a way as to align the four LFs at the bright
end.  Vertical shifts of the M30, M92, and NGC 6397 LFs were made to
bring them into alignment with the M15 LF, according to a
least-squares algorithm, in the magnitude intervals $4<$ \MI\ $<7$ and
$5<$ \MV\ $<7.5$.  The shift adopted for NGC 6397 was determined from
the \hst\ measurements alone, but is further reinforced by the good
match between the \hst\ and ground-based measurements.

With the LFs aligned in this way, it can be seen that the \Ieight-band
LFs of all clusters have a similar shape in the range $4\leq$ \MI\ $\leq
7$.  At fainter magnitudes the LFs of M15, M30, and M92 continue to
track one another closely, remaining similar over the entire range
sampled, from the the turnoff down to \MI $\simeq 10$, or ${\cal M}\sim
0.13$\msun.  The NGC 6397 LF, by contrast, diverges downward for \MI\
\simgr\ 7, or \MV\ \simgr\ 8 (\mass\ \simless\ 0.5\msun), dropping below
those of M15, M30, and M92 by as much as a factor of 2.5.  A similar
relative deficit of faint stars in NGC 6397 vs.\ M15, M30, and M92 is
visible in the \Vfive\ LFs in Fig.\ 2b.

Had we chosen to align the NGC 6397 LF with the others at the faint
instead of the bright end, the result would have been a relative
excess of bright stars in NGC 6397.  However, the range of magnitudes
over which the NGC 6397 LF would be well-matched to the other three
LFs would in this case be somewhat smaller than when they are aligned
at the bright end.
Nevertheless, the present observations do not in themselves clearly
distinguish between an excess of high-mass stars or a deficit of
low-mass stars in this cluster, relative to the other three.  But for
reasons that will be discussed in Section 5 below, we will hereafter
refer to the difference in the LFs as a deficit of faint stars in NGC
6397.

The difference between the NGC 6397 LF and the nearly identical LFs of
M30, M15, and M92 is apparent only when they are compared over a large
range of magnitudes.  De Marchi \& Paresce (1995) have
asserted---incorrectly, as it turns out---that NGC 6397 and M15 have
very similar LFs.  This was the result of the shortness of the
interval over which they compared their LFs---only \MI\ $\simeq
6.5$--10.1.  Our \hst\ LFs for NGC 6397 and M15 span a larger range of
magnitudes, allowing a comparison from \MI\ $\simeq 3.5$--10.5.  
These
LF measurements are in reasonable agreement with those of De Marchi \&
Paresce for M15, and with Paresce, De Marchi \& Romaniello (1995) for
NGC 6397, for the ranges in which we overlap with them; but the larger
magnitude range that we measure shows that the two LFs have different
shapes:\ that they match each other at the bright end, and that there
is a marked deficiency of faint stars in NGC 6397 relative to M15 (as
well as M30 and M92).

Before we can interpret the results of the LF comparison in Fig.\ 2, we
must determine what corrections, if any, are required to convert these
observed local LFs to global LFs.  Our multimass model of NGC 6397 (King,
Sosin, \& Cool 1995) shows that the LF in its envelope differs little from
the global LF, as mass-segregation effects, while strong, are largely
confined to the small central regions.  
(The radial changes are shown in more detail in Figure 4 of King 1996.)
To first order, these results are
applicable to M15 and M30 as well, since all three clusters have collapsed
cores (Djorgovski \& King 1986) and similar surface-brightness profiles,
and since the fields analyzed here are out in the envelopes in all three
cases.  
More specifically, even with the small rescaling that is needed in
fitting the NGC 6397 model to M15 and M30, their global MFs remain very
close to the local MFs in the fields that we observed, differing nowhere
by more than one or two tenths in the logarithm.
As for M92, we have calculated multi-mass models of clusters of
similar central concentration.  We find that whereas local mass
functions at the center or very far out in the envelope differ
considerably from the global one, at the intermediate radius at which we
observed the local mass function is fortuitously quite close to the
global one, with differences that again nowhere exceed one or two tenths in
the logarithm.
Thus in all four clusters our observed LFs are effectively global, and
the large deficiency of faint stars in NGC
6397 is very unlikely to be due to differences between the
local and global LFs.

\section{Mass Functions}

Deep color--magnitude arrays also contain information about the
low-mass-star content of clusters.  But in order to extract this
information, a transformation from luminosity to mass is required.  As
many authors have emphasized, such transformations remain uncertain for
low-mass stars, especially for very-low-mass stars of low metal
abundance.  The transformation from an LF to an MF depends directly on
the slope of the mass--luminosity relation (MLR), and different
calculations of stellar models yield different slopes, particularly for
the lowest-mass hydrogen-burning stars.  To get a sense of the range of
results that different MLRs produce, we have converted our LFs into MFs
using several of those available in the literature (Bergbush and
VandenBerg 1992, Baraffe \etal\ 1995, D'Antona \& Mazzitelli 1995,
Alexander \etal\ 1996).  Caution must still be exercised in interpreting
the resulting MFs, however, given the underlying problem of the paucity
of observational constraints on any of these relations.

Of the existing models, we find that those of Alexander
\etal\ (1996) reproduce the location of our main sequence in the CMD
of NGC 6397 unusually well, as shown in Fig.\ 3.  Not only is the
overall fit of the isochrone from the model that has the metallicity of
NGC 6397 satisfactory, from just below the turnoff to the very bottom of
the observed main sequence, but also the distance modulus $(m-M)_V=12.4$
and the reddening $E(V-I)=0.19$ that result from the fit are in
agreement, within the errors, with the distance and reddening of NGC
6397 found by Piotto \etal\ (1996) from an average of different methods.
This result is somewhat reassuring, in the sense that we can expect that
if the models are able to predict the luminosities and temperatures, and
also the corrections from the theoretical to the observational CMD
plane, they should then be able also to predict the transformation from
masses to absolute magnitudes, i.e., the MLR that we need. This does not
mean that the MLR from the Alexander \etal\ (1996) models is the
``correct'' one; it is simply the best one presently available, and the
cautionary remarks we made at the beginning of this section should still
be heeded.  The main sequences of M15, M30, and M92 all have very
similar morphologies to that of NGC 6397.  As a result, what has been
said for the isochrone fitting of NGC 6397 also applies to the other
three metal-poor clusters.  To conclude, neither the D'Antona and
Mazzitelli (1995) nor the Baraffe \etal\ (1995) isochrones reproduce the
observed sequence of NGC 6397 as well as that of Alexander \etal\
However, for the sake of comparison, and to give an idea of the possible
range of uncertainty, we will nevertheless use the MLR of D'Antona \&
Mazzitelli along with that of Alexander \etal\ in what follows.  The
former MLR produces the steepest MFs of any we tried, while the latter
produces the shallowest MFs.

In Fig.\ 4 we compare the MFs derived from the \Ieight\ LF of NGC 6397
using both MLRs.  The results are similar over most of the main
sequence, from ${\cal M} \simeq0.8$--0.15\msun.  But for the
lowest-mass stars (\mass\ $\simeq 0.15$--0.10\msun), the Alexander
\etal\ MLR produces a MF that falls significantly below that produced
by the D'Antona \&\ Mazzitelli relation.  This is a consequence of the
steeper slope that the MLR of D'Antona \&\ Mazzitelli has in this
region.

Neither MF is particularly well fit by a power law over the entire
mass range.  For purposes of comparison with previous work, however,
we have determined best-fit power laws for both MFs for masses below
0.4\msun.  They are $x = -0.1$ and $x = 0.6$ for the MFs derived from
Alexander \etal\ and D'Antona \&\ Mazzitelli MLRs, respectively (where
for the Salpeter law $x = 1.35$).  Neither is as steep as the $x =
0.9$ slope determined by Fahlman \etal\ (1989) for this cluster.
However that steep slope was derived from a correspondingly steep LF
that disagrees with two independent sets of HST observations (see
CPK). The discrepancy could be the result of inadequate correction for
field stars in the ground-based study, a problem that we have already
described as being critical for NGC 6397.

In Fig.\ 5 we compare the MFs derived from the \Ieight\ LF of M30,
again using both the D'Antona \& Mazzitelli and the Alexander \etal\
MLRs.  Similar results are obtained for M15 and M92, since their LFs
are so similar to that of M30.  Again, neither MF is well fit by a
single power law over the entire length of main sequence that is
sampled.  If the upper end of the main sequence is ignored, the MF
derived using the D'Antona \&\ Mazzitelli MLR can be reasonably well
fit by a power law with slope $x \sim 1$ for stars with \mass\ $<$
0.4\msun.  The MF obtained using the Alexander \etal\ MLR is similar,
but somewhat less well fit by a power law, even in this restricted
range of masses.  It shows some hint of flattening at the
very-low-mass end, which is not seen in the MF that the D'Antona \&\
Mazzitelli MLR produces.  We emphasize that of the MLRs that we have
tried, its ability to match our CMD makes the Alexander \etal\ MLR the
one that is probably most reliable, though a direct comparison with
observed masses will ultimately be needed.

\section {Discussion}

The primary finding of this study is that the main-sequence LFs and
MFs of M15, M30, and M92 are almost as similar as they can be, while
that of NGC 6397 is distinctly different.  How are we to interpret
this result?

The resemblance between M15, M30, and M92 would tend to suggest that
these three clusters were born with similar MFs that have changed
little since, or changed in similar ways.  A scenario in which they
were born different and have evolved to become the same seems too
contrived.  That NGC 6397 has a different LF may simply be an accident
of birth.  However, in view of the similarity between the other three
clusters, it is interesting to ask whether the differences in NGC 6397
could be understood if all four clusters started out with similar mass
functions, and evolved in different ways.  
Viewing the difference in the NGC 6397 LF as a relative deficit of
low-mass stars, the answer---qualitatively, at least---is yes.

Of the more than 100 globular clusters for which physical parameters
are given by Djorgovski (1993), NGC 6397 has the shortest central
relaxation time; those of M15, M30, and M92 range from one to three
orders of magnitude longer.  Also, among the 26 Galactic orbits given
by Dauphole \etal\ (1996), NGC 6397 has one that is among the most
vulnerable to tidal shocks; it oscillates rapidly through a dense part
of the Galactic plane, only a few kiloparsecs from the Galactic
center.  By contrast, M15 has an orbit that subjects it very little to
tidal shocks:\ it is much farther from the center, and slower moving
(Dauphole \etal\ 1996).  M92 may be an intermediate case; its orbit,
which carries it quite close to the Galactic center at times, would
make it vulnerable to shocks.  However, this case is quite different
from that of NGC 6397.  Passages through the dense central part of the
Galactic disk are much less frequent for M92, because its orbit is
much more extended, and therefore has a much longer period.  Moreover,
the central relaxation time of M92 is so long that the possibility
exists that the cluster has lost low-mass stars from its periphery but
replenishes them so slowly from its inner parts that the loss no
longer continues.  However, the possibility that M92 has been affected
by tidal shocks cannot be excluded. And we note that the faintest part
of its LF does appear to fall somewhat below those of M15 and M30,
though the difference is only at the 2-sigma level and is also
affected by how we fit the curves together.

As no measurement of the proper motion of M30 exists, its Galactic
orbit is poorly known; but its present distance from the Galactic
plane and its radial velocity suggest that it has an orbit that is
much less vulnerable to tidal shocks than is that of NGC 6397.  The
radial velocity of M30 is $-$186 km/s, at
$(l,b)=(27^\circ,-47^\circ)$.  If its proper motion were zero, this
radial velocity would indicate a speed, with respect to the Galactic
center, of 240 km/s, sufficiently larger than circular to carry the
cluster considerably farther from the Galactic center than its present
distance of 6.8 kpc.  While it is not beyond the realm of possibility
that the unknown proper motion might reduce the magnitude of the
cluster's Galactocentric speed, the great majority of possible values
of the proper motion would increase it.  Thus M30 is most probably in
an orbit much more like the low-shock orbit of M15 than the high-shock
orbit of NGC 6397.

Both evaporation and losses via tidal shocking could be expected to
have resulted in a depletion of low-mass stars in NGC 6397 relative to
M15, M30, and M92.  Moreover, the two effects should work in concert,
with relaxation feeding low-mass stars to the cluster envelope as fast
as the tidal shocks remove them.  We emphasize, however, that the
rates of these effects are poorly known; detailed calculations of both
the evaporation rate and the effects of tidal shocks on this cluster
would be valuable.
(Aguilar, Hut, \& Ostriker [1988] carried out a study of these effects,
showing their importance; but they treated the orbit of each cluster in
a statistical way, for lack of knowledge---at that time---of actual
orbits.  
A more recent study by Gnedin \& Ostriker [1997], although more
sophisticated in its approach, chose to ignore known orbits, ``to
maintain homogeneity of the sample''; they also treated orbits
statistically.) 

We further find that differences between existing MLRs for
very-low-mass, metal-poor main-sequence stars are sufficient to
produce noticeably different MFs from the same LF.  The differences we
find are greatest for NGC 6397, for which the LF that we measure
reaches to the faintest absolute magnitudes.  It remains unclear
exactly how its MF behaves at the very-low-mass end.  But the MF
slopes that we find using any of the existing MLRs are shallower than
the $x = 1$ slope for which an integration of the total mass down to
\mass\ = 0 would diverge.  Thus, NGC 6397 appears not to contain a
significant amount of mass in very-low-mass stars.  At the same time,
even the shallowest MF we obtained for NGC 6397 shows no sign of
dropping, to the limit to which we can identify the main sequence.  As
for the bottom of the hydrogen-burning MS, it is clear that we have not
yet reached it, as that region would manifest itself as a sharp drop-off
in the LF.

M15, M30, and M92 all contain a larger fractional population of
very-low-mass main-sequence stars than NGC 6397.  For these three
clusters, the best-fit power law for the MF at the low-mass end is
close to the divergent slope of $x = 1$ for stars with \mass\ $<$
0.4\msun.  However, none of the MLRs we tried produced MFs for these
clusters that are as steep as those reported for a small sample of
other clusters by Richer \etal\ (1991).  Furthermore, since most of
the MLRs we tried produced MFs that begin to flatten out near the
low-mass limit of our measurements, we conclude that the MFs probably
do not rise so fast as to make the low-mass end dominant.  It is
important to emphasize, however, that globular-cluster MFs will remain
uncertain until the mass--luminosity relation for very-low-mass,
low-metallicity stars is better determined.  Fortunately, such
uncertainties do not affect comparisons between clusters with similar
metallicities, and much progress in that area can be anticipated in
the near future.

\acknowledgments

We thank F.\ D'Antona, V.\ Castellani, S.\ Cassisi, and E.\ Brocato for
providing isochrones in advance of publication, D.\ VandenBerg for
providing a computer-readable version of his isochrones, and P.\ Stetson
for his generosity with software.  We also note that the M15 data used
here exists as a result of the foresight of E.\ Groth in setting up a
parallel-observing program for HST.  This work was supported by NASA
Grant NAG5-1607.  GP acknowledges partial support by the Consiglio
Nazionale delle Ricerche and by the Agenzia Spaziale Italiana.

\newpage

\noindent{Fig.~1---} ($V$, $V-I$) color--magnitude diagrams of M15 (NGC
7078), M30 (NGC 7099), M92 (NGC 6341), and NGC 6397.  Note that two
different $V$ filters were used.

\noindent{Fig.~2a---} \MI\ LFs for NGC 6397, M15, M30, and M92. 
The LF for NGC 6397 has been extended up to the turnoff using
ground-based data.  Note the similarity between the \MI\ LFs for M15,
M30, and M92, and the significantly shallower slope of the LF for NGC
6397 relative to those of the other three clusters.

\noindent{Fig.~2b---} $M_V$ LFs for NGC 6397, M15, M30, and M92. 
The LF for NGC 6397 has been extended up to the turnoff using
ground-based data.  As in Fig.\ 2a, note the similarity between the $M_V$
LFs for M15, M30, and M92, and the significantly shallower slope of the
LF for NGC 6397 relative to those of the other three clusters.

\noindent{Fig.~3---} Comparison of our NGC 6397 CMD with theoretical
curves from Alexander \etal\ (1996).

\noindent{Fig.~4---} Mass functions for NGC 6397, converted
from the \Ieight\ LFs shown in Fig.\ 1, using the MLRs from Alexander
\etal\ (1996) (solid line) and from D'Antona \& Mazzitelli (1995)
(dotted line).  Three specific masses referred to in the text are
marked with arrows.

\noindent{Fig.~5---} Same as Fig.~4, but for M30 (NGC 7099).

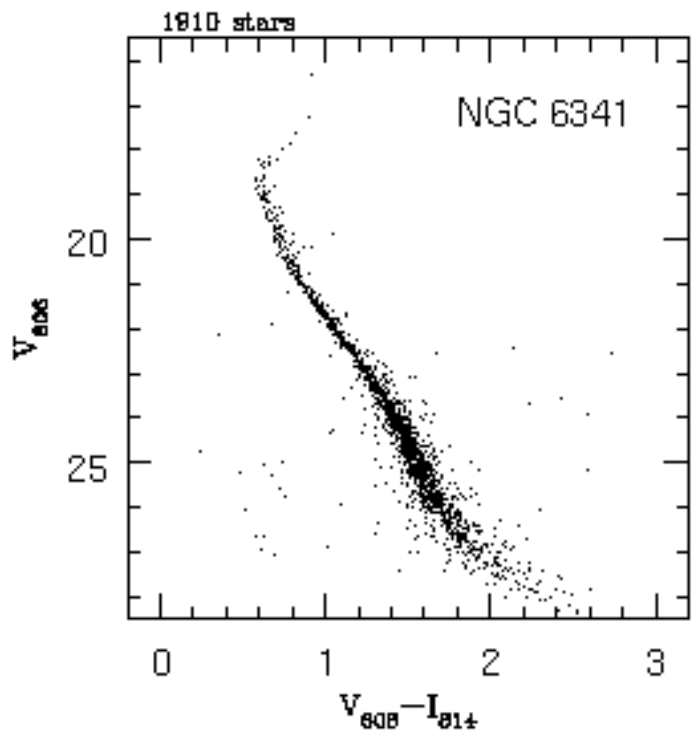
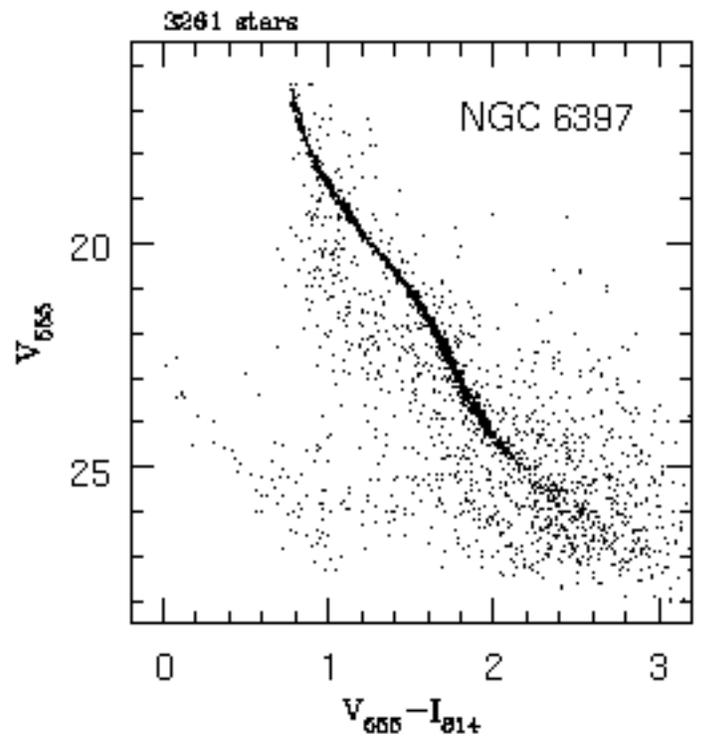
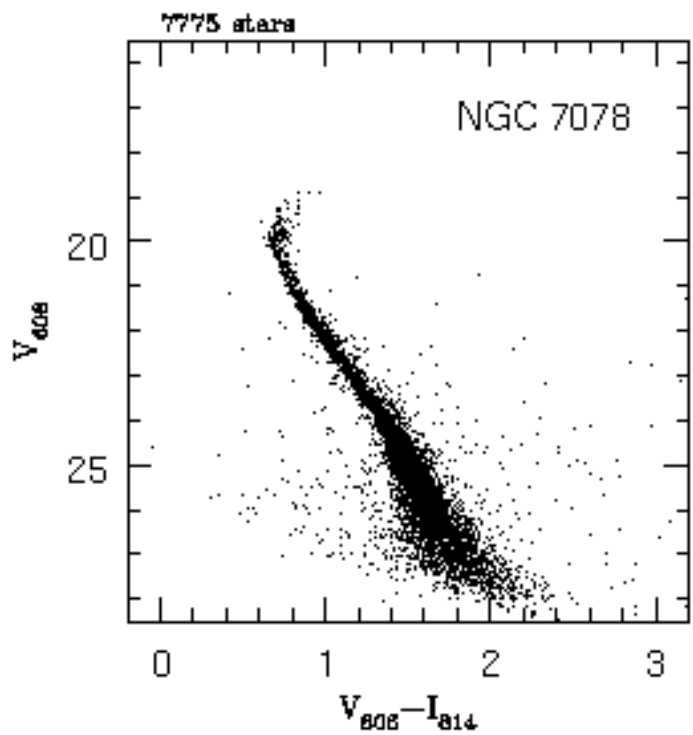
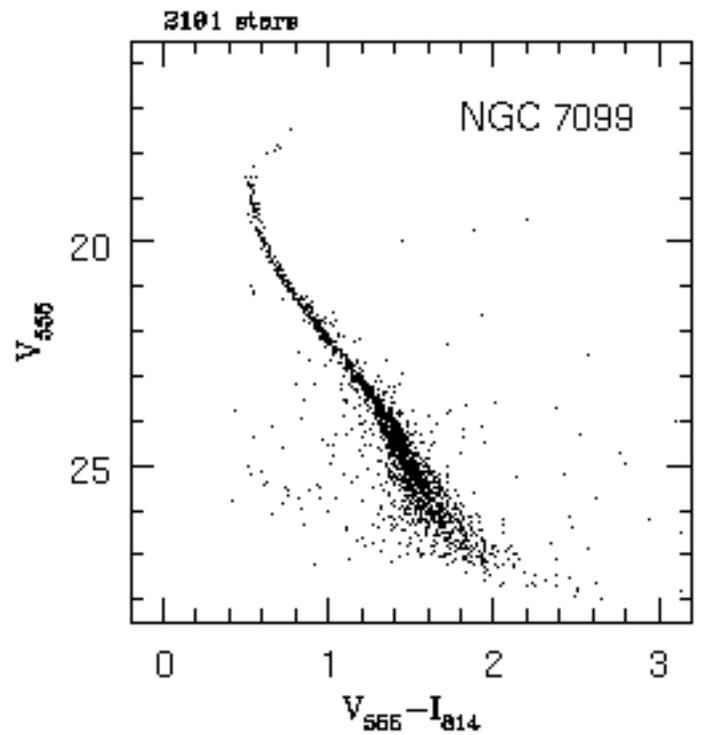
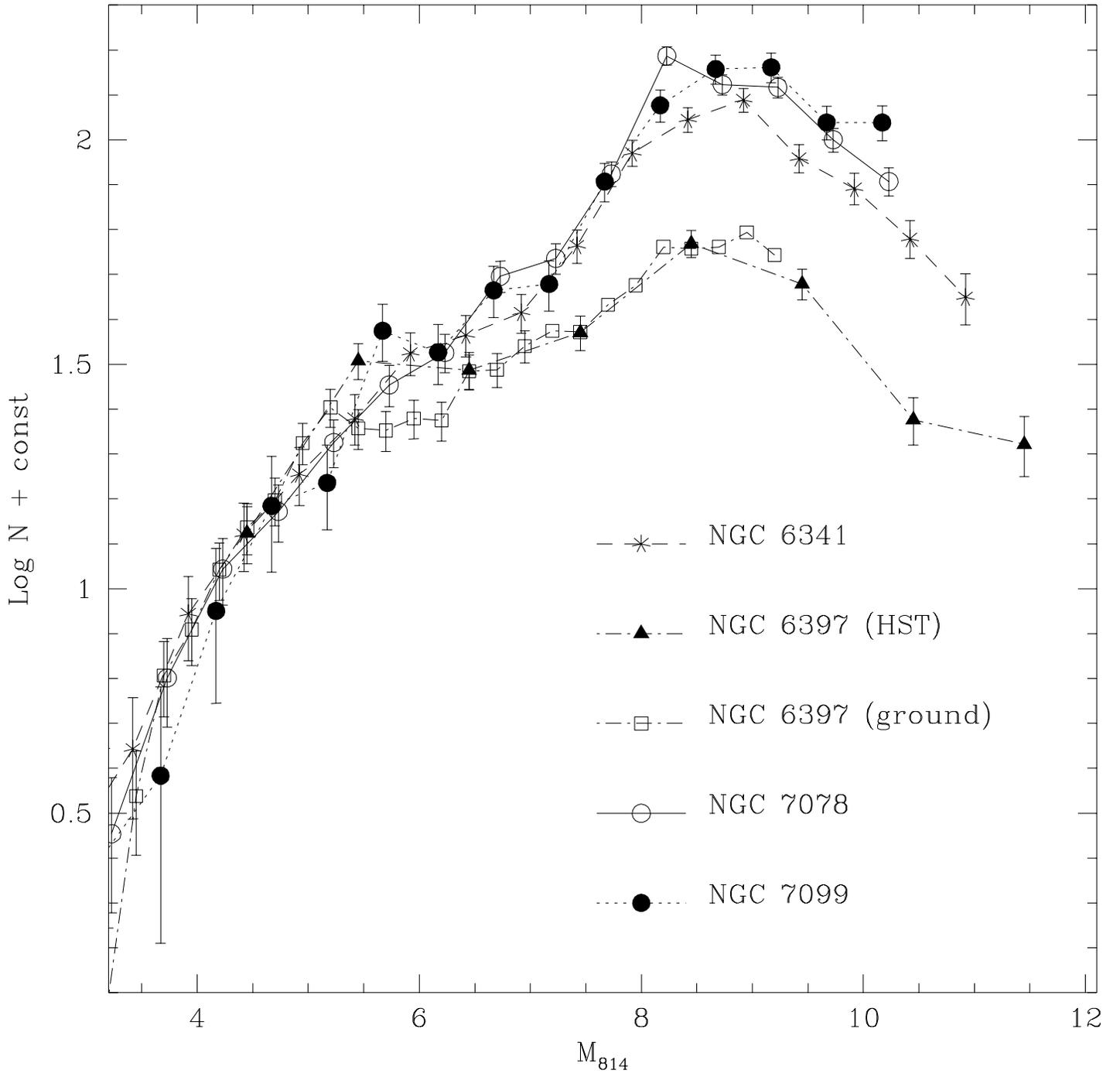
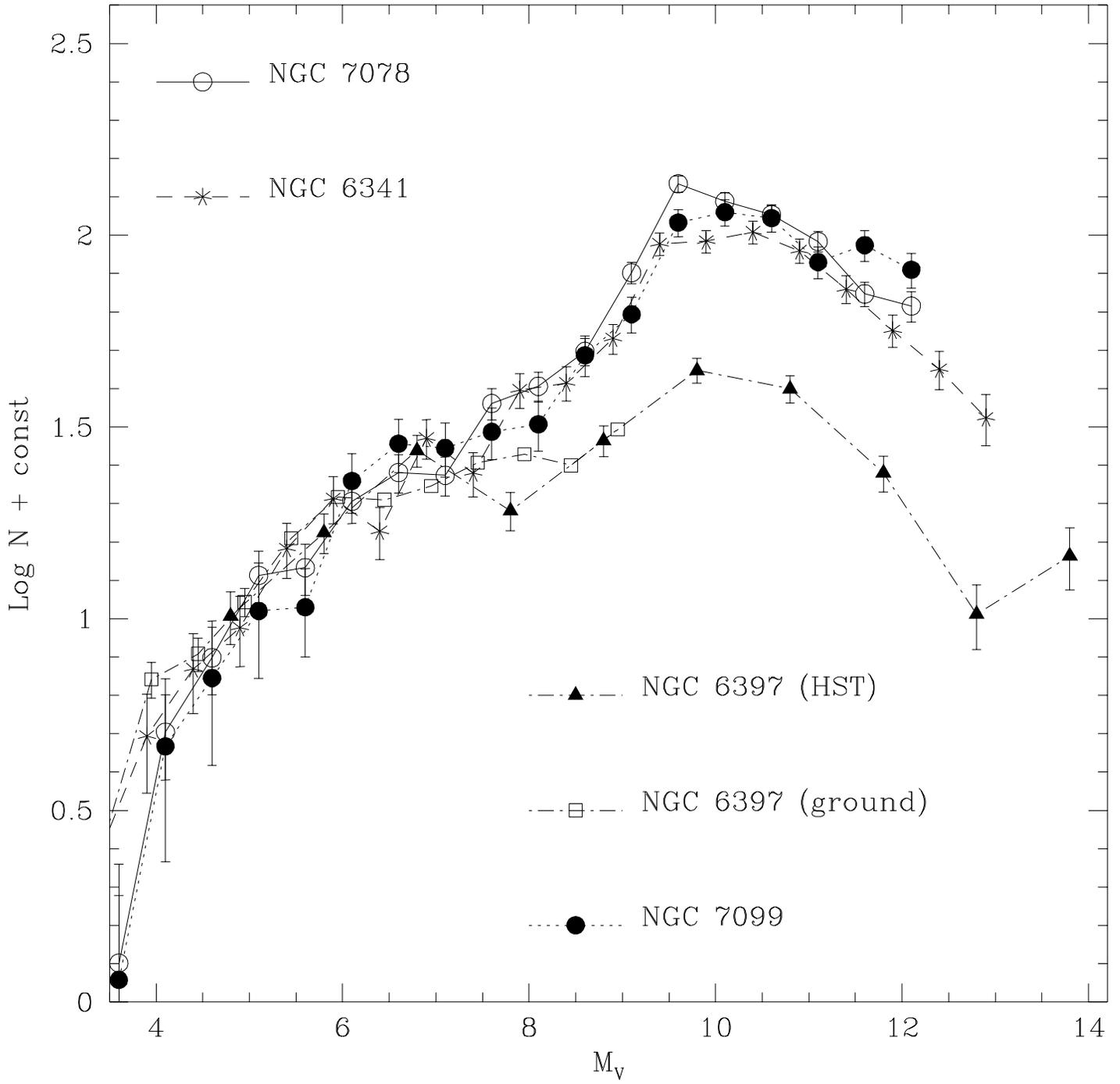
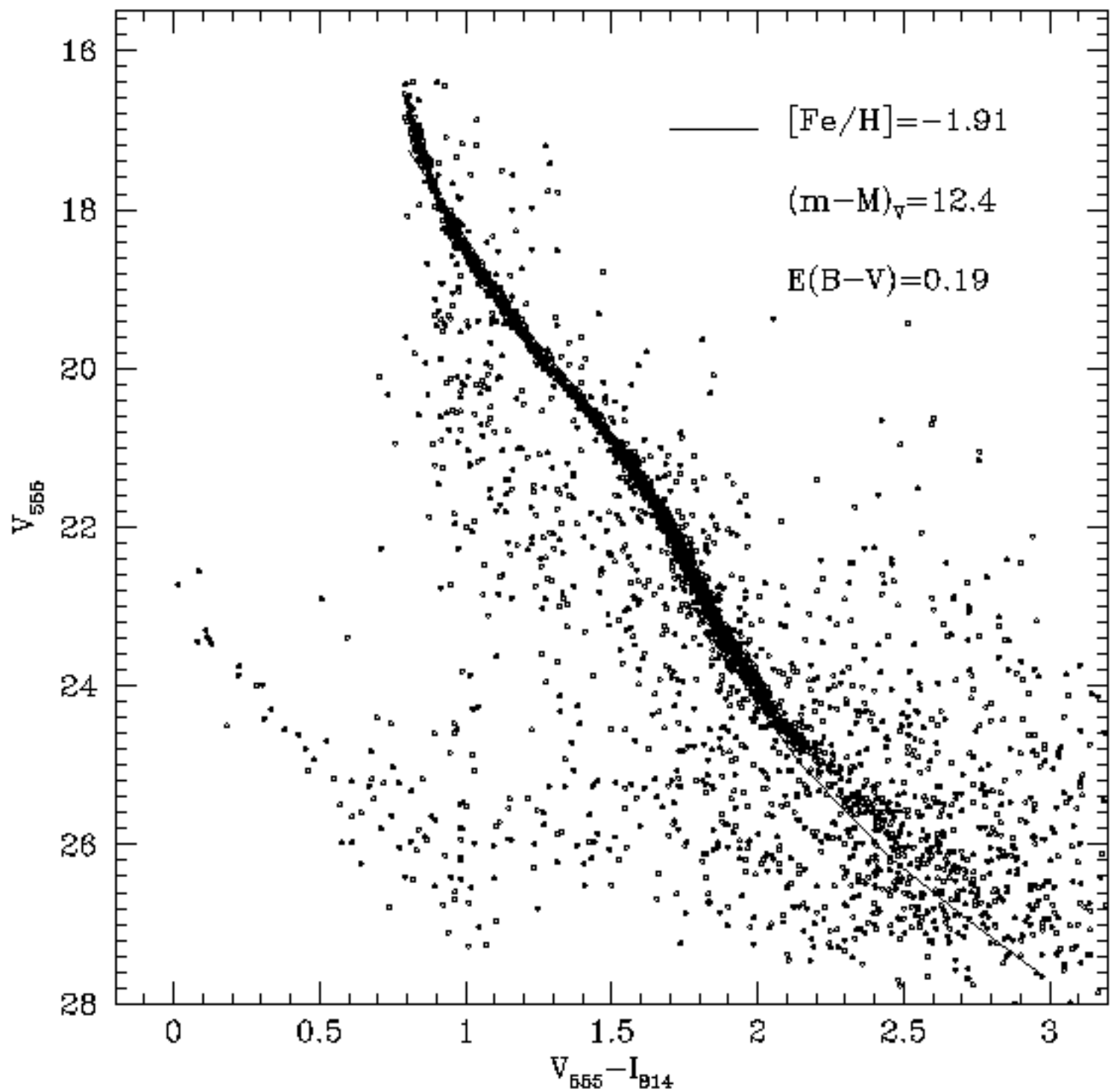
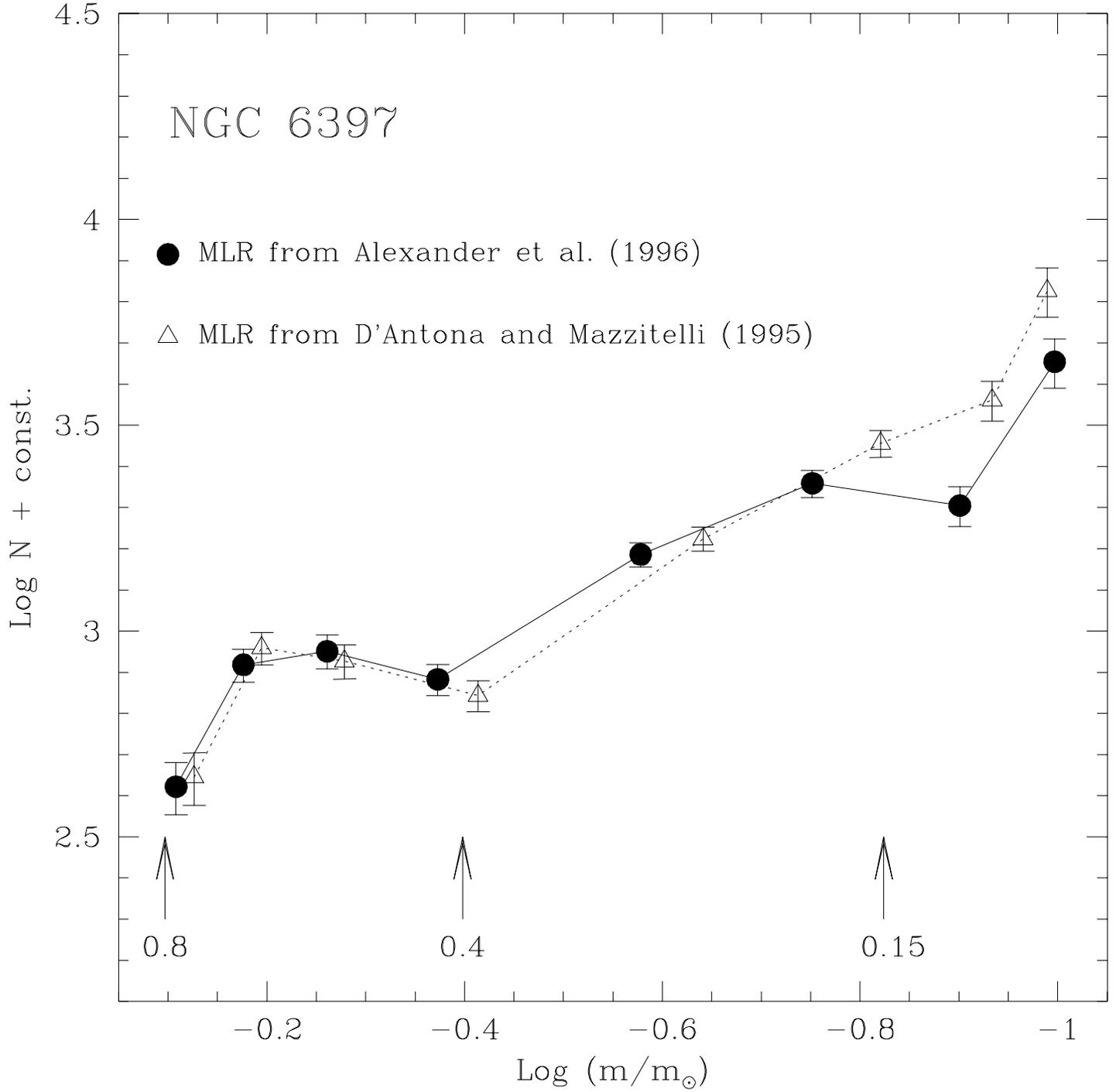
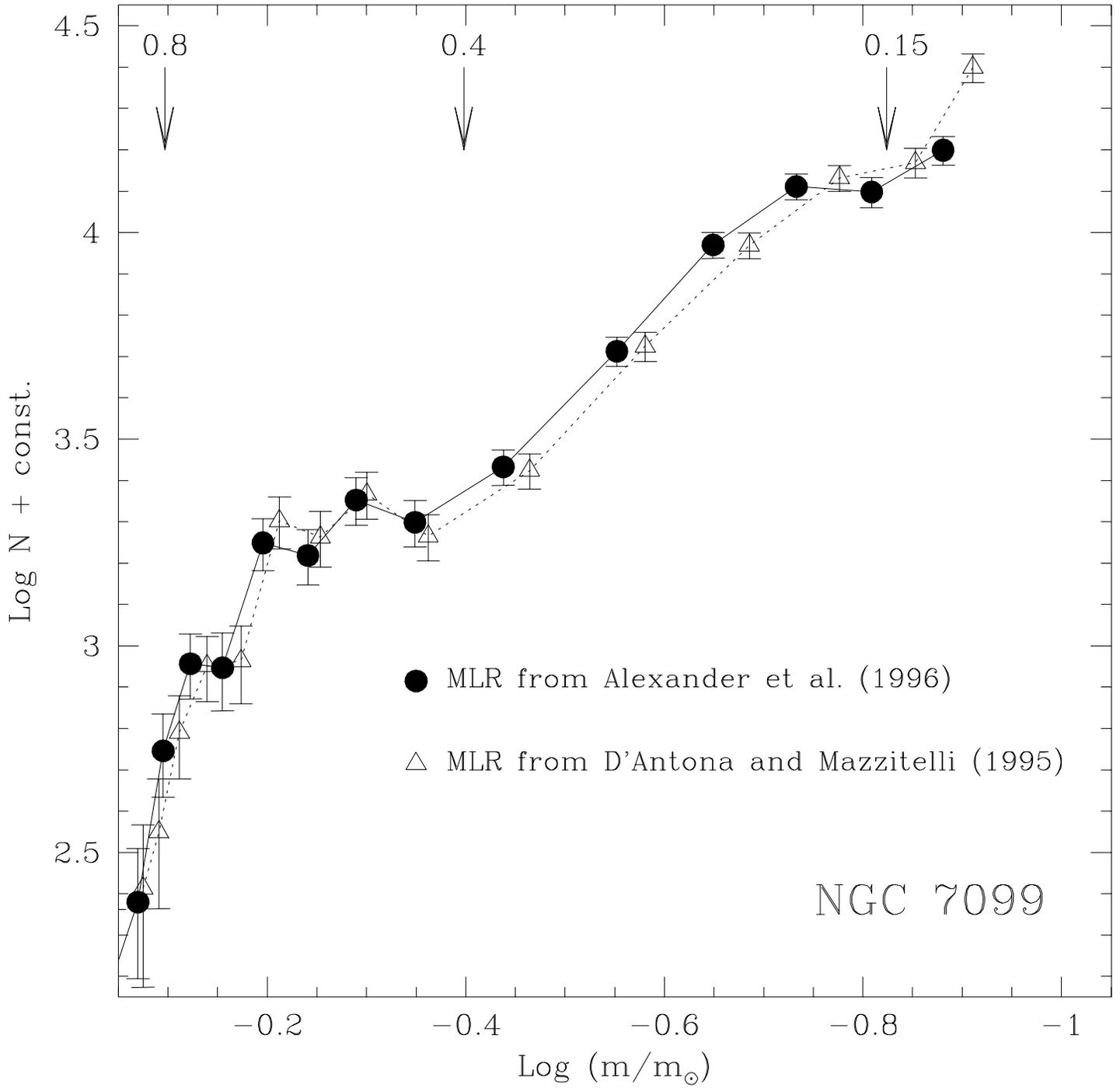
\end{document}